\documentclass[journal]{IEEEtran}
\ifCLASSINFOpdf \else \fi

\usepackage{cite}
\usepackage{amsmath,amssymb,amsfonts}
\usepackage{algorithmic}
\usepackage{gensymb}
\usepackage{graphicx}
\usepackage{textcomp}
\usepackage{rotating}
\usepackage{multirow}
\usepackage{tabularx}
\usepackage{threeparttable}
\usepackage{xcolor}
\usepackage{amsmath}
\usepackage[caption=false]{subfig}
\usepackage{capt-of}
\usepackage{booktabs}
\mathchardef\mhyphen="2D % Define a "math hyphen"
\newcommand{\rpm}{\sbox0{$1$}\sbox2{$\scriptstyle\pm$}
  \raise\dimexpr(\ht0-\ht2)/2\relax\box2 }
\usepackage[
colorlinks=true,
urlcolor=blue
]{hyperref}
\def\BibTeX{{\rm B\kern-.05em{\sc i\kern-.025em b}\kern-.08em
    T\kern-.1667em\lower.7ex\hbox{E}\kern-.125emX}}
\begin{document}
\title{Attention-aware non-rigid image registration for accelerated MR imaging}
\author{Aya Ghoul, Jiazhen Pan, Andreas Lingg, Jens Kübler, Patrick Krumm, Kerstin Hammernik, Daniel Rueckert (Fellow IEEE), Sergios Gatidis, Thomas Küstner (Member, IEEE)
\thanks{This work was supported by the Deutsche Forschungsgemeinschaft (DFG, German Research Foundation) under Germany’s Excellence Strategy – EXC 2064/1 – Project number 390727645.}
\thanks{Aya Ghoul, Sergios Gatidis and Thomas Küstner are with the University Hospital of Tuebingen, Department of Diagnostic and Interventional Radiology, Medical Image and Data Analysis (MIDAS.lab), Tübingen, Germany (e-mail: \{aya.ghoul; sergios.gatidis; thomas.kuestner@med.uni-tuebingen.de)}
\thanks{Jiazhen Pan and Daniel Rueckert are with Klinikum Rechts der Isar, Technical University of Munich, Munich, Germany (e-mail: jiazhen.pan; daniel.rueckert@tum.de)}
\thanks{Kerstin Hammernik is with the School of Computation, Information and Technology, Technical University of Munich, Munich, Germany (e-mail: k.hammernik@tum.de)}
\thanks{Andreas Lingg, Jens Kübler and Patrick Krumm are with the University Hospital of Tuebingen, Department of Diagnostic and Interventional Radiology, Tübingen, Germany (e-mail: andreas.lingg; jens.kuebler; patrick.krumm@med.uni-tuebingen.de) }
\thanks{Daniel Rueckert is also with the School of Computation, Information and Technology, Technical University of Munich, Munich, Germany and the Department of Computing, Imperial College London, London, United Kingdom.}
}

\maketitle

\begin{abstract}
 
Accurate motion estimation at high acceleration factors enables rapid motion-compensated reconstruction in Magnetic Resonance imaging (MRI) without compromising the diagnostic image quality. In this work, we introduce an attention-aware deep learning-based framework that can perform non-rigid pairwise registration for fully sampled and accelerated MRI. We extract local visual representations to build similarity maps between the registered image pairs at multiple resolution levels and additionally leverage long-range contextual information using a transformer-based module to alleviate ambiguities in the presence of artifacts caused by undersampling. We combine local and global dependencies to perform simultaneous coarse and fine motion estimation. The proposed method was evaluated on in-house acquired fully sampled and accelerated data of 101 patients and 62 healthy subjects undergoing cardiac and thoracic MRI. The impact of motion estimation accuracy on the downstream task of motion-compensated reconstruction was analyzed. We demonstrate that our model derives reliable and consistent motion fields across different sampling trajectories (Cartesian and radial) and acceleration factors of up to 16x for cardiac motion and 30x for respiratory motion and achieves superior image quality in motion-compensated reconstruction qualitatively and quantitatively compared to conventional and recent deep learning-based approaches. The code is publicly available at https://github.com/lab-midas/GMARAFT.

\end{abstract}


\section{Introduction}
\label{sec:introduction}
\IEEEPARstart{M}{agnetic} resonance imaging (MRI) is commonly used for disease diagnosis as it provides high soft tissue contrast and structural information. However, scan times are relatively long to ensure high spatial and temporal resolution for the different imaging contrasts and sequences. This makes imaging in the body trunk region prone to cardiac and respiratory motion artifacts that translate into blurring or ghosting and hence deteriorate image quality. Handling of motion by multiple breath-holds or triggered imaging allows for scanning within quiescent periods\cite{edelman1991coronary, higgins1984magnetic, hatabu1999fast}. Nonetheless, these scanning protocols remain lengthy and may cause patient discomfort. Alternatively, signal averaging techniques correct for repetitive motion by segmenting k-space data into different motion states and exploiting correlations between them or warping the images into a single state \cite{odille2008generalized}. Other methods rely on motion-robust data acquisition schemes including radial trajectories\cite{feng2014golden}, PROPELLER\cite{pipe1999motion} and spiral sampling\cite{liao1997reduction}. Such trajectories are often more motion-tolerant than Cartesian trajectories due to the innate averaging of the oversampled k-space center. Accelerated imaging addresses motion implicitly by shortening acquisitions either using parallel imaging \cite{kellman2008fully,kellman2009high,griswold2002generalized,pruessmann1999sense} or random undersampling with model-based sparsity \cite{osher2005iterative,chaari2011wavelet,block2007undersampled,liu2013adaptive} and/or low rank \cite{lingala2011accelerated, knoll2019assessment, bustin2019five, kuestner20193d} reconstructions or more recently relying on deep learning-based methods \cite{schlemper2017deep, kustner2020cinenet, sandino2021accelerating}. Despite enabling more efficient scans, high acceleration factors trade off the signal-to-noise ratio for speed. Hence, the reconstructions may suffer from residual aliasing artifacts and reduction of edge sharpness that compromise the image resolution and diagnostic value. Motion information can be leveraged to simultaneously improve the spatial resolution and reduce motion artifacts either by enhancing the sparsity and/or low-rankness of the sparse representations \cite{ Hansen2012retrospective, kuestner2021fully, Liu2020dynamic} or by embedding motion into a motion-compensated reconstruction \cite{batchelor2005matrix, usman2013motion, schmidt2011nonrigid}. 

Different registration algorithms have been applied to estimate motion from motion-resolved images by identifying a smooth transformation that maps one moving image to another fixed image without additional distortions. However, the intricate non-rigid movement and the underlying through-plane motion of 2D imaging make finding precise spatial and temporal correspondences challenging. Different deformation models have been introduced in the literature including free-form B-splines\cite{rueckert1999nonrigid}, elastic models \cite{bajcsy1989multiresolution}, diffusion-based models \cite{bro1996fast}, diffeomorphic transforms \cite{avants2008symmetric} and optical flow-based models \cite{horn1981optical}. Such conventional methods formulate the motion estimation task as a variational problem solved iteratively by optimizing an energy function. For each image pair, the hyperparameters are carefully fine-tuned and the optimization problem is solved anew to ensure optimal results. Hence, motion estimation becomes tedious and slow in practice. 

Recent learning-based approaches have been presented to learn generalizable registrations and achieved the accuracy of traditional methods while being significantly faster\cite{cao2017deformable,rohe2017svf, kustner2021lapnet, vos2017end, morales2019implementation, balakrishnan2019voxelmorph}. A global function is learned during training and evaluated on unseen images replacing the need for the costly subject-specific optimization. Recent works adapted self-supervised learning, where an image similarity metric is optimized while enforcing spatial smoothness\cite{vos2017end, morales2019implementation, balakrishnan2019voxelmorph}. Others demonstrated state-of-the-art performance by encouraging the model to learn more meaningful representations through an additional auxiliary task like aligning segmentation masks\cite{qin2018joint} or solving non-rigid registration and image reconstruction jointly\cite{qi2021end, yang2022end, pan2022learning}. Furthermore, different training loss formulations have been proposed including cyclic constraints to enhance inverse consistency \cite{kim2021cyclemorph} and enforcing diffeomorphism to guarantee topology preservation \cite{mok2020fast}. Other methods investigated the use of advanced hyperparameter search mechanisms to facilitate finding optimal hyperparameter values, such as hypernetworks \cite{hoopes2021hypermorph} and conditional instance normalization \cite{mok2021conditional}.

Most of the proposed methods are convolutional neural networks (CNN) that apply dense filters to extract local image features. Although CNNs provide detailed high-resolution spatial information, they have difficulties learning explicit long-range dependencies due to the intrinsic locality of the convolution operation\cite{luo2016understanding}. Theoretically, we can increase the sizes of the kernels to cover larger displacements. However, the number of trainable parameters will increase drastically and, hence, the computational cost to train and test the model. To overcome this limitation, recent works adopted vanilla transformer architectures\cite{shi2022xmorpher, zhang2021learning}. Their innate self-attention mechanism determines which parts of the image are relevant based on high-level contextual representations. Such insight is valuable for gaining a deeper understanding of the spatial relationships between the registered images. However, training transformers is computationally expensive and results may suffer from limited localization due to insufficient low-level details\cite{wang2018non}. Other methods have proposed hybrid architectures by complementing CNNs with transformers\cite{chen2021vit,chen2022transmorph} to combine low-level features with global context. These methods focused on estimating motion from fully sampled images. However, fully sampled and motion-resolved k-spaces are expensive to collect and even unavailable for some applications where the scan duration is limited. 

To leverage the full potential of spatial-temporal information sharing across motion cycles, non-rigid image registration should be ideally performed as early as possible during data acquisition, necessitating its capability to handle data from accelerated MRI scans. This enables reducing the scan duration while maintaining high spatial and temporal resolution comparable to breath-held and segmented acquisitions. In this work, we propose the following contributions: (1) We present a novel self-supervised deep learning framework for pairwise non-rigid image registration of fully sampled and accelerated MRI data. Our method performs concurrently coarse and fine motion estimation by benefiting from self-attention to preserve global contextual information. High-level information mitigates the effect of undersampling artifacts by contributing to learning better representations of the underlying image structures. (2) In contrast to earlier works that adapted mainly patch-based processing for non-rigid image registration of undersampled images, we process the entire image while keeping a low computational overhead and without additional pre- and post-processing steps. This reduces inconsistencies in the estimated motion field and allows for capturing the coordinated and interdependent motion of distant anatomical structures more efficiently. (3) We additionally leverage the image features of a denoised image to alleviate the impact of aliasing artifacts. We inject the smoothed image representation into the self-attention operator to permit better aggregation of motion within the extracted spatial information. We tested our approach throughout acceleration factors of up to 16x for cardiac motion and 30x for respiratory motion in different sampling trajectories (Cartesian and radial) in a cohort of 101 patients and 62 healthy subjects. The successful use of our motion estimation framework within the downstream task of motion-compensated reconstruction is demonstrated. Our method preserved high image fidelity and outperformed other conventional and deep learning-based methods.

\begin{figure*}[h]
\centerline{\includegraphics[width=\textwidth]{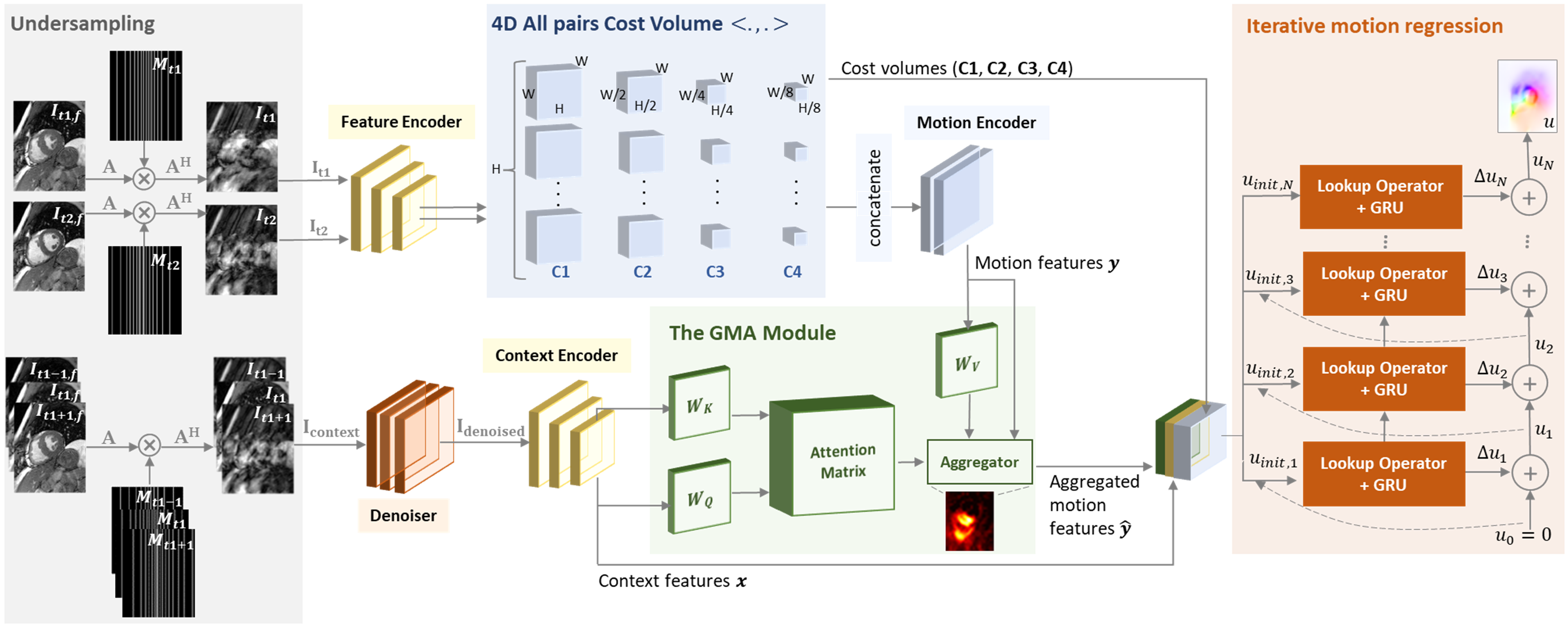}}
\caption{Illustration of the proposed framework for non-rigid image registration with exemplary Cartesian VISTA undersampling. Local feature maps are extracted from the undersampled magnitude images $I_{t1}$ and $I_{t2}$ obtained by undersampling the fully sampled images $I_{t1,f}$ and $I_{t2,f}$. $A$ and $A^{H}$ refer to the multi-coil forward and backward operators used for the Cartesian undersampling. Visual correspondences are calculated as the inner product between all pairs of the encoded features of size $W\times H$, pooled to multiple levels to obtain multi-scale tensors $C_{1\text{-}4}$ and then encoded into the motion features. The denoiser smoothes the context images and outputs the denoised images that are encoded into the context features. High-level context is obtained using self-attention embedded in the Global Motion Aggregation module (GMA). $W_\text{Q}$, $W_\text{K}$ and $W_\text{V}$ are projection heads of the queries, keys and values. The GRU decodes for $N$ iterations the flow initialization $u_{init,i}=u_{i-1}$ and the collected feature maps into update directions $\Delta u_{1\text{-}N}$ to refine the motion estimates $u_{1\text{-}N}$.}
\label{architecture}
\end{figure*}

\section{Methods}

\subsection{Proposed Registration Network}

The proposed network architecture\footnote{Our code is available online at \url{https://github.com/lab-midas/GMARAFT}.} is illustrated in Figure \ref{architecture} and is inspired by RAFT \cite{teed2020raft} and its extension \cite{jiang2021learning}. Initially, the fixed and the moving image are encoded separately using the Feature encoder that starts with one convolutional layer with $64$ filters and a $7\times7$ kernel size using a $2$-stride, followed by $6$ consecutive residual blocks and another convolutional layer with $256$ filters, a $1\times1$ kernel size and $1$-stride. Each residual unit includes two sequential units of convolution, group normalization\cite{wu2018group} and rectified linear unit (ReLU) activation operations. The residual units have respectively $32$, $32$, $64$, $64$, $128$ and $128$ filters with $3\times3$ kernels. Within the third and fifth residual blocks, downsampling operations are integrated by applying a stride of $2$. Conversely, the remaining units use a stride of $1$, maintaining the input spatial dimensions. Concurrently, the fixed image and its two neighboring frames are denoised using a ResNet-15 network in the second branch \cite{he2016deep}. Subsequently, they are fed into another encoder, sharing an identical architecture with the Feature encoder, to extract the context features. The cross-image pixel correspondences are computed as the pairwise dot-product between all pairs of feature vectors of size ($H \times W \times D$) extracted from the fixed and moving images of size ($8H \times 8W \times 1$). For each feature vector extracted from the fixed image, we store a 4D $(W \times H \times W \times H)$ cost volume to generate a 2D response map for each pixel of the moving image feature map \cite{scharstein2002taxonomy}. The cost volume scales a quadratic computational complexity $\{\mathcal{O}(HW)^2D\}$ with respect to the feature maps. We apply volume cascading by using average pooling with different kernels of sizes $1$, $2$, $4$, and $8$ to build multi-scale cost volumes and allow for capturing large and small displacements. These multi-scale volumes are subsequently concatenated along the channel dimension and encoded into 2D motion features using two consecutive convolution layers with $128$ and $256$ filters and respectively $1\times1$ and $3\times3$ kernels. The aggregated motion features are derived from the context and the motion feature maps using the attention-based Global Motion Aggregation (GMA) module\cite{jiang2021learning}. We cast the registration task as an optimization and operate on the learned features imitating the update steps of solving a first-order descent algorithm for better generalization and outlier rejection to determine a solution. The used update operator performs lookups on the 4D multi-scale cost volumes and uses 2D convolutional gated recurrent unit (GRU) \cite{cho2014properties} to find the motion estimation iteratively. We define a local search neighborhood around the estimated match of each pixel in the moving frame based on a motion initialization $u_{init}$, following \cite{teed2020raft}. $u_{init}$ is initialized as zeros at all pixels, during the first iteration. Depending on the determined search grids, the lookup operator generates a feature map by indexing from all levels of the 4D cost volumes and storing individual local explicit representations of the displacements corresponding to each pixel. We decode the local motion features indexed by the lookup operator, the aggregated motion features and the context features to adjust the GRU weights. At each iteration $i$, a residual update direction $\Delta{u}_{i}$ is deduced and added to the previous estimation to obtain the current motion estimation $u_{i}$ and the next motion initialization $u_{init,i+1}$ to update the local cost volumes for the next iteration, as follows:
\begin{equation}
u_{init,i+1} = u_{i} = u_{init,i} + \Delta{u}_{i}
\end{equation}
The update operator performs iterative motion regression such that the sequence converges to a fixed point. The network outputs a sequence of $N$ dense motion estimates, denoted as $\{u_1, ...,u_N\}$, corresponding to $N$ iterations, where each estimate is refined through a consecutive iteration of the motion estimation process.  

\subsubsection{The GMA Module}
Anatomical structures have generally homogeneous or interdependent motion. Hence, if we find the pixels that belong to every anatomical entity by looking at similar appearances in the fixed frame, we can identify pixels with similar and related motion patterns. This information will help the model decipher ambiguities at boundary levels. Transformers can better comprehend such overall spatial correspondences thanks to their self-attention mechanism\cite{raghu2021vision}. They process the entire input sequence and capture long-term dependencies, that are essential for learning hierarchical global attributes. We aim to enhance the motion features by incorporating global context through instilling self-similarities between pixels of the fixed image in the hidden representation of the multi-scale motion information. Thus, we project the context features into queries and keys forming hidden representations of the fixed image features that store self-similarity in appearance feature space. On the other hand, the motion feature map is projected using a separate learned projector to form the values. For the feature vector $i$ of the context features $x \in \mathbb{R}^{HW\times D_c}$, the corresponding aggregated motion feature vector $\hat{y_i}$ is obtained from the motion features $y \in \mathbb{R}^{HW\times D_m}$ and the context features $x$, as follows \cite{jiang2021learning}:
\begin{equation}
\hat{y_i}=y+\alpha\sum_{j=1}^{HW}\operatorname{Softmax}\left(\frac {(W_\text{Q}x_i)^T \cdot W_\text{K}x_j}{\sqrt{D_c}}\right)W_\text{V}y_j
\end{equation}
where $W_\text{Q}$, $W_\text{K} \in \mathbb{R}^{D_c\times D_c}$ and $W_\text{V} \in \mathbb{R}^{D_m\times D_m}$ are learnable projection heads for queries, keys and values. $\alpha$ is a learned weight for the influence of attention on the prediction. $D_c$ and $D_m$ refer to the channel dimension of the context and the motion feature maps.

\subsubsection{The Denoiser Module}

We use a modified ResNet-15 \cite{he2016deep} to denoise the undersampled artifact-reduced images. The module has simplified residual blocks with removed batch normalization layers. The input is first fed to a convolutional layer that has three $3\times3$ filters followed by a ReLU activation and is forwarded subsequently to four successive residual blocks. Each block has three convolutional layers with $32$ filters of respectively $3\times3$, $3\times3$ and $1\times1$ kernel sizes.

\subsection{Datasets}
\label{sec:data}

Multi-slice 2D cardiac cine images were acquired on a $1.5$T MRI scanner (MAGNETOM Aera, Siemens Healthcare, Erlangen, Germany). Short axis cine acquisitions were performed with a 2D balanced steady-state free precession (bSSFP) sequence in $65$ patients ($45\rpm17$ years, $24$ female) with suspected cardiovascular diseases and $38$ healthy subjects ($32\rpm6$ years, $14$ female). Short-axis slices were acquired in $8$ breath-holds of $15$s each. Other imaging parameters were: TE=$1.06$ms, TR=$2.12$ms, slice thickness=$8$mm, spatial resolution=$1.9\times1.9\text{mm}^2$ with a matrix size in the range of $176 \times 132$ to $192 \times 180$, temporal resolution=$40$ms, $25$ reconstructed cardiac phases, flip angle=$52^\circ$ and bandwidth=$915$Hz/px. We performed retrospective undersampling to create the training and test data. We used a variable density incoherent spatiotemporal acquisition (VISTA)\cite{ahmad2015variable} undersampling for the Cartesian trajectory to mask k-spaces. An inverse Fourier operation and coil sensitivity maps form the backward operation $A^{H}$ which is used to get the undersampled (zero-filled) coil-combined images from the masked multi-coil k-spaces. Coil sensitivity maps are estimated from the acquired auto-calibration signal data\cite{uecker2014espirit}. For the non-Cartesian trajectory, we used a golden angle radial sampling and applied an adjoint Non-Uniform Fast Fourier Transform (NUFFT) to get the undersampled images. We accounted for the inhomogeneous distribution with a density compensation function (DCF) and regridded the sampled k-space into a Cartesian grid using an interpolation kernel following \cite{winkelmann2006optimal}. The studied models were trained using one undersampling strategy with accelerations ranging from $R\!=\!1$ (fully sampled) to $R\!=\!16$. 

We used a second dataset to evaluate the performance of our model on respiratory motion. 3D motion-resolved k-space data was obtained from $36$ patients ($60\rpm9$ years, $21$ female) with suspected liver or lung metastases and $24$ healthy subjects ($32\rpm8$ years, $10$ female) on a 3T PET/MR (Biograph mMR, Siemens Healthcare, Erlangen, Germany). A 3D T1 weighted spoiled gradient echo sequence was used with a continuous variable-density Poisson Disc Cartesian undersampling. Other imaging parameters were TE=$1.23$ms, TR=$2.60$ms, resolution=$1.9 \times 1.9 \times 1.9$ \text{mm}$^3$ covering a field-of-view of $500 \times 500 \times 360 \text{mm}^3$ with a matrix size $256 \times 256 \times 144$ (FH $\times$ LR $\times$ AP), flip angle=$7^\circ$ and bandwidth=$890$Hz/px. The data was retrospectively gated based on self-navigation into distinct $4$ respiratory bins reordered from end-expiratory to end-inspiratory with a Gaussian view-sharing amongst neighboring gates\cite{kuestner2021fully, kustner2017self}. We used a variable-density Poisson-Disc (vdPD) undersampling for training and testing with accelerations that ranged from $R\!=\!1$ to $R\!=\!30$. 

All subjects were prospectively recruited and gave written consent for participation in the trial and publication of their imaging data and results. The studies were approved by the local ethics committee (426/2021BO1, 721/2012BO1).

\subsection{Training and Implementation Details}

\begin{figure*}[h]
\centerline{\includegraphics[width=\textwidth]{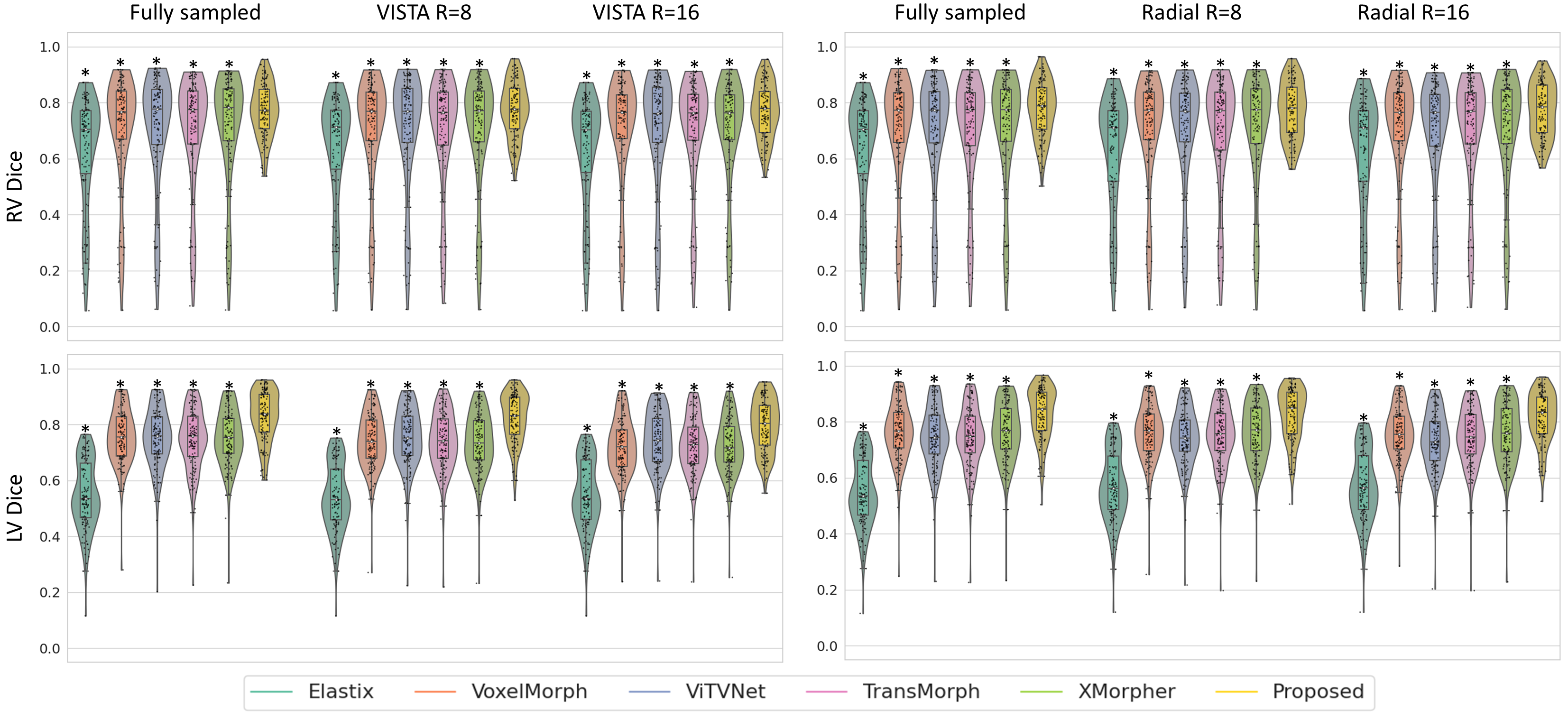}}
\caption{Quantitative evaluation between registered segmentation masks and manually annotated masks over the fully sampled and retrospectively undersampled acquisitions with VISTA (Cartesian) and radial (non-Cartesian) undersampling for $R=8$ and $R=16$ accelerations using the Dice loss for left ventricle (LV) and right ventricle (RV) with motion estimation using Elastix\cite{klein2009elastix}, VoxelMorph\cite{balakrishnan2019voxelmorph}, Vit-V-Net\cite{chen2021vit}, TransMorph\cite{chen2022transmorph}, XMorpher\cite{shi2022xmorpher} and the proposed method. Our model outperforms the other works on observed structures in terms of Dice score. The symbol $*$ denotes a statistically significant difference, indicated by a p-value of less than 0.05 when compared to our method.}
\label{DSC}
\end{figure*}

We used 2D coil-combined magnitude images as the network's input. Training data was created by taking image pairs from $53$ patients and $29$ healthy subjects for the cardiac dataset and from $30$ patients and $19$ healthy subjects for the respiratory dataset. The remaining subjects were not included in training and used only for testing. Each 4D image  (x,  y,  slice,  time frame) was split up slice-by-slice and then frame-by-frame to create 2D input pairs $(x,  y)$. One training candidate was defined as a pair of image frames within the same slice at two-time points of the motion cycle. In total, we created $531250 / 32448$ image pairs for training and $109375 / 5760$ image pairs for testing in the cardiac/respiratory dataset per acceleration factor.  Each 2D input image was center-cropped to reduce memory requirements and allow batch processing and then scaled in the range $[-1, 1]$. During testing, zero padding was applied to the motion estimate to match the original shape of the image. 

The network outputs a sequence of $N$ dense motion estimates $\{u_1, ...,u_N\} $ corresponding to $N$ decoding iterations. We used a self-supervised loss function following \cite{vos2017end, balakrishnan2019voxelmorph}. The model was updated using three loss functions. The photometric loss $\mathcal{L}_{photo}$ penalizes differences between the fully sampled fixed image $I_{t1,f}$ and the motion-corrected image obtained by warping the fully sampled moving image $I_{t2,f}$ with the estimated motion $u_i$ at iteration $i$ using a spatial transformation function $T$ that applies bilinear interpolation:
\begin{equation}
\mathcal{L}_{photo,i} = ||I_{t1,f} - T(I_{t2,f}, u_i)||_1 
\label{eq:photo_loss}
\end{equation}
The smoothness loss $\mathcal{L}_{smooth}$ ensures local spatial smoothed motion and is computed with an anisotropic diffusion regularizer on the spatial gradients of the motion estimate:
\begin{equation}
\mathcal{L}_{smooth,i} = ||\nabla u_i||_1
\label{eq:smooth}
\end{equation}
where the Jacobian is computed by approximating spatial gradients along the image directions using forward finite derivatives differences in both directions to calculate the partial derivatives. Additionally, we used the mean-squared error as a denoising loss $\mathcal{L}_{denoiser}$ between the denoiser output $I_{denoised}$ and the fully sampled images $I_{context,f}$ corresponding to the context branch input:
\begin{equation}
\mathcal{L}_{denoiser} = ||I_{denoised} - I_{context,f}||_2
\end{equation}
The final loss $\mathcal{L}_{final}$ was computed as the combination of the denoiser loss and the weighted cumulative sum of the photometric and smoothness loss with an exponentially increasing coefficient $\delta$ at each iteration on the fully sampled images: 
\begin{equation}
\mathcal{L}_{final} = \delta \mathcal{L}_{denoiser} + \sum_{i=1}^{N} \beta^{N-i} (\mathcal{L}_{photo,i} + \gamma \mathcal{L}_{smooth,i}) 
\end{equation}
where the hyper-parameters $N$, the denoising regularization parameter $\delta$, the iteration weight $\beta$ and the smoothness regularization parameter $\gamma$ were determined using a grid search with the goal of minimizing the photometric loss of selected validation subjects from the test set between the warped image, obtained by transforming the fully sampled moving image with the estimated motion, and the corresponding fixed image. The parameters are set to $N$=$12$, $\delta$=$0.5$, $\beta$=$0.8$ and $\gamma$=$0.2$. We used an AdamW \cite{loshchilov2017decoupled} optimizer with a batch size of 8 and a one-cycle learning rate scheduler with the highest learning rate set to $5 \times 10^{-4}$. The model was implemented using PyTorch and has $_{\widetilde{~}}5.95$ million trainable parameters. Training required $36$ hours on an NVIDIA V100 GPU.

\subsection{Motion-compensated Reconstruction}
\label{sec:recon}
We used the motion estimates of the investigated registration in the downstream task of motion-compensated reconstruction based on a generalized matrix formulation approach\cite{batchelor2005matrix} to reconstruct the whole imaged volume. The motion estimation is embedded in the forward model to leverage temporal redundancy in the reconstruction. Given the motion-compensated image $I_t$ corresponding to the temporal frame $t$ to be reconstructed from $\tau = \rpm T$ neighboring frames of the frame $t$, the motion-corrupted k-space $m_t$ is formulated as: 
\begin{equation}
m_t = \sum_{\tau=\mhyphen T}^{T} A_{\tau}U_{t\rightarrow \tau} I_t = \sum_{\tau=\mhyphen T}^{T} M_{\tau}FSU_{t\rightarrow \tau} I_t
\label{eq:moco}
\end{equation}
where $A_{\tau}=M_{\tau}FS$ denotes the forward MR operation that includes the sampling operator $M_{\tau}$, the coil sensitivities $S$ and Fourier transform $F$. The coil sensitivities are estimated as described in section \ref{sec:data} and remain constant for all frames. $U_{t\rightarrow \tau}$ is the cardiac motion matrix that transforms the neighboring cardiac frames $\tau \in [\mhyphen T,T]$ of the reconstructed images $I$ to the current time frame $t$. This reconstruction problem is solved using iterative SENSE\cite{pruessmann2001advances}. 

\section{Experiments}

\subsection{Motion Estimation and Baselines}
We conducted comprehensive experiments on cardiac data for fully sampled and accelerated acquisitions using data from $21$ test subjects excluded from the training subjects ($12$ patients and $9$ healthy subjects). The VISTA undersampling (Cartesian) and radial golden-angle (non-Cartesian) trajectory were investigated to infer the robustness of the model on cardiac motion estimation against different trajectories and artifact types. High accelerations were addressed to permit accelerated imaging with whole-heart coverage within a single breath-hold. We also studied the performance of our model on respiratory motion of $8$ test subjects ($6$ patients and $2$ healthy subjects) for fully sampled and accelerated data obtained with a Cartesian mask to examine how well our model performs on another non-rigid motion pattern. 

We compared the motion estimate obtained at iteration $N=12$ by our model with the results of prior works that demonstrated state-of-the-art registration performance. Elastix\cite{klein2009elastix}, a 3D parametric B-spline-based registration, served as the conventional baseline. We fine-tuned the hyperparameters of Elastix with the objective of eliminating distortion while minimizing the photometric loss between the fixed and transformed images. We used the Parzen Window Mutual Information \cite{mattes2001nonrigid} as a similarity metric with $4$ multi-resolution levels and $32$ histogram bins. Each resolution underwent optimization over $500$ iterations, with a random coordinate image sampler extracting $2048$ points from a randomly selected sample region in each iteration. We compared also our method to other deep learning registrations that have different architectures (CNN, transformer and hybrid transformer-CNN models), including VoxelMorph (CNN)\cite{balakrishnan2019voxelmorph}, Vit-V-Net (hybrid)\cite{chen2021vit}, TransMorph (hybrid)\cite{chen2022transmorph}, XMorpher (transformer)\cite{shi2022xmorpher}. All baseline algorithms were executed following the source code provided by the authors and trained on our databases. We conducted a grid search to fine-tune the parameter settings such that the photometric loss of selected validation subjects from the test set between the warped image and the fixed image was minimized. The hyperparameters included the training similarity loss $\mathcal{L}_{sim}$, the smoothness regularization parameter $\gamma$, the weight decay $\lambda$ and the learning rate $\eta$. The used parameters are summarized in Table \ref{tab_hyperparams}. The performance remained consistent whether the training was conducted on fully sampled data exclusively or incorporated additionally undersampled data for both the fully sampled and accelerated validation cases. Other parameters adhered to the default implementation, as proposed in the respective publications. Table \ref{tab_FLOP} shows the number of trainable parameters and the inference floating-point operations (FLOP), necessary for processing a single pair of input images, using the baselines and the proposed method.

\begin{table}[h]
\caption{Hyperparameters of the comparative methods.}
\label{table}
\centering
\begin{tabular}{p{1.2cm} p{3.25cm} p{0.5cm} p{0.5cm} p{1.2cm}}
 \hline
Model & $\mathcal{L}_{sim}$& $\gamma$ & $\lambda$ &  $\eta$\\
 \hline
VoxelMorph	&mean squared error & $0.04$ & $10^{-4}$ & $10^{-4}$  \\
Vit-V-Net	&mean squared error & $0.02$ & $10^{-4}$ & $2\times10^{-4}$  \\
TransMorph	&mean squared error & $0.04$ & $10^{-4}$ & $5\times 10^{-4}$  \\
XMorpher	&normalized cross-correlation & $0.06$ & $10^{-3}$ & $5\times 10^{-4}$  \\

 \hline
\end{tabular}
\label{tab_hyperparams}
\end{table}

\begin{table}[h]
\caption{The parameters and FLOP comparison of the studied registration methods.}
\label{table}
\centering
\begin{tabular}{lcc}
 \hline
Model & Params (M) & FLOP (G) \\
 \hline
VoxelMorph & $0.29$  &	$73.39$\\
Vit-V-Net  & $110.62$&  $54.99$\\
TransMorph & $46.78$ &  $114.43$\\
XMorpher   & $20.51$ &  $155.44$\\
Proposed   & $5.95$  &  $30.68$\\

 \hline
\end{tabular}
\label{tab_FLOP}
\end{table}

\begin{table*}[]
\centering

\caption{Quantitative comparisons of Cartesian (top) and radial (bottom) motion-compensated reconstructions to ground-truth reconstructions for fully sampled, $R\!=\!8$ and $R\!=\!16$ accelerated data using SSIM, PSNR, NRMSE and HFEN. Results for frame-to-frame registration using $\mathcal{L}_{photo}$ and regularity of deformation fields measured with the percentage of pixels with a non-positive Jacobian determinant $|J_{\phi}|$. Average and standard deviations are reported.}
\begin{threeparttable}
\label{tab_comp_vista}
\begin{tabular}{clllllll}
\hline 
\multicolumn{2}{l}{\textbf{Cartesian VISTA}} & Elastix & VoxelMorph & Vit-V-Net & TransMorph & XMorpher & Proposed \\ \hline
 
\multirow{6}{*}{fully sampled} 

& SSIM &$0.82\rpm0.07^*$ &$0.83\rpm0.05^*$ &$0.82\rpm0.05^*$ &$0.73\rpm0.04^*$ &$0.75\rpm0.10^*$ &$\mathbf{0.85\rpm0.05}$   \\
 & PSNR &$39.17\rpm3.38^*$ &$40.30\rpm3.24^*$ &$39.45\rpm3.44^*$ &$35.64\rpm2.86^*$ &$36.80\rpm4.80^*$ &$\mathbf{40.97\rpm3.65}$    \\
 & NRMSE &$0.18\rpm0.05$ &$0.19\rpm0.08$ &$0.20\rpm0.08^*$ &$0.27\rpm0.08^*$ &$0.27\rpm0.11^*$ &$\mathbf{0.18\rpm0.08}$   \\
 & HFEN  ($\times 10^{\text{\scriptsize -2}}$) &$2.03\rpm0.71^*$ &$2.14\rpm0.66^*$ &$2.28\rpm0.69^*$ &$3.14\rpm0.61^*$ &$2.76\rpm0.87^*$ &$\mathbf{1.87\rpm0.77}$  \\
 & $\mathcal{L}_{photo}$($\times10^{\text{\scriptsize -3}}$) &$8.04\rpm4.46^*$&$5.91\rpm1.54^*$&$6.89\rpm2.98^*$&$11.35\rpm3.73^*$&$7.56\rpm4.54^*$&$\mathbf{5.36\rpm2.86}$  \\
 & $\perthousand$ of $|J_{\phi}| \leq 0$ &$0.01\rpm0.03^*$ &$\mathbf{0.01\rpm0.01^*}$ &$2.12\rpm1.21^*$ &$4.16\rpm1.68^*$ &$3.15\rpm8.50^*$ &$0.24\rpm0.07$     \\ 
 
\\
 
\multirow{6}{*}{$R=8$} 

& SSIM &$0.68\rpm0.08^*$ & $\mathbf{0.78\rpm0.06}$ & $0.61\rpm0.06^*$ & $0.62\rpm0.05^*$ & $0.66\rpm0.12^*$ & $\mathbf{0.78\rpm0.06}$ \\
 & PSNR &$34.95\rpm3.27^*$ &$37.78\rpm3.07$ &$33.23\rpm2.84^*$ &$32.85\rpm2.37^*$ &$34.63\rpm4.41^*$ &$\mathbf{37.82\rpm3.19}$  \\
 & NRMSE &$0.51\rpm0.05^*$ &$0.23\rpm0.07$ &$0.31\rpm0.06^*$ &$0.32\rpm0.07^*$ &$0.31\rpm0.12^*$ &$\mathbf{0.21\rpm0.07}$ \\
 & HFEN  ($\times 10^{\text{\scriptsize -2}}$) &$5.18\rpm0.65^*$ & $2.31\rpm0.71^*$ &$3.87\rpm0.66^*$ &$4.21\rpm0.74^*$ &$2.91\rpm0.92^*$ &$\mathbf{2.23\rpm0.76}$ \\
 & $\mathcal{L}_{photo}$($\times10^{\text{\scriptsize -3}}$) &$8.37\rpm4.56^*$&$6.31\rpm1.06^*$&$15.91\rpm5.38^*$&$13.42\rpm3.86^*$&$6.87\rpm4.15^*$&$\mathbf{5.82\rpm2.21}$  \\
 & $\perthousand$ of $|J_{\phi}| \leq 0$ &$1.72\rpm1.02^*$ &$\mathbf{< 0.01^*}$ &$3.51\rpm1.14^*$ &$4.82\rpm1.63^*$ &$1.30\rpm4.82^*$ &$0.08\rpm0.02$  \\ \\
 
\multirow{6}{*}{$R=16$} 

& SSIM  &$0.51\rpm0.08^*$ &$0.72\rpm0.06^*$ &$0.49\rpm0.06^*$ &$0.53\rpm0.05^*$ &$0.61\rpm0.12^*$ &$\mathbf{0.74\rpm0.07}$    \\ 

 & PSNR &$32.72\rpm3.15^*$ &$35.65\rpm3.06^*$ &$31.04\rpm2.74^*$ &$31.09\rpm2.28^*$ &$33.48\rpm3.98^*$ &$\mathbf{36.04\rpm3.08}$    \\
 
 & NRMSE &$0.38\rpm0.05^*$ &$0.24\rpm0.07$ &$0.38\rpm0.06^*$ &$0.37\rpm0.07^*$ &$0.33\rpm0.12^*$ &$\mathbf{0.23\rpm0.07}$   \\
 
 & HFEN  ($\times 10^{\text{\scriptsize -2}}$) &$4.11\rpm0.84^*$ &$2.83\rpm0.78^*$ &$5.13\rpm0.98^*$ &$4.97\rpm0.88^*$ &$3.41\rpm0.96^*$ &$\mathbf{2.48\rpm0.86}$   \\
 & $\mathcal{L}_{photo}$($\times10^{\text{\scriptsize -3}}$) &$8.06\rpm4.53^*$&$6.68\rpm0.94^*$&$18.07\rpm6.02^*$&$13.98\rpm4.02^*$&$6.83\rpm4.29^*$&$\mathbf{5.95\rpm2.17}$  \\
 & $\perthousand$ of $|J_{\phi}| \leq 0$ &$2.37\rpm1.39^*$ &$\mathbf{< 0.01^*}$ &$4.71\rpm1.41^*$ &$4.81\rpm1.68^*$ &$1.20\rpm4.64^*$ &$0.03\rpm0.04$ \\ \hline
\\[-0.9em]
\hline
\multicolumn{2}{l}{\textbf{non-Cartesian radial}} & Elastix & VoxelMorph & \multicolumn{1}{l}{Vit-V-Net} & TransMorph & XMorpher & Proposed \\ \hline
\multirow{6}{*}{fully sampled} 
& SSIM  &$0.82\rpm0.07^*$ &$0.83\rpm0.05^*$ &$0.82\rpm0.05^*$ &$0.80\rpm0.05^*$ &$0.72\rpm0.12^*$ &$\mathbf{0.85\rpm0.05}$  \\
 & PSNR &$39.17\rpm3.38^*$ &$40.41\rpm3.28^*$ &$40.68\rpm3.60^*$ &$37.81\rpm2.88^*$ &$35.40\rpm4.71^*$ &$\mathbf{41.16\rpm3.23}$   \\
 & NRMSE &$0.18\rpm0.05$ &$0.19\rpm0.08^*$ &$0.20\rpm0.08^*$ &$0.21\rpm0.08^*$ &$0.30\rpm0.14^*$ &$\mathbf{0.17\rpm0.08}$    \\
 & HFEN  ($\times 10^{\text{\scriptsize -2}}$) &$2.03\rpm0.71^*$ &$2.08\rpm0.66^*$ &$2.32\rpm0.71^*$ &$2.38\rpm0.64^*$ &$3.12\rpm1.12^*$ &$\mathbf{1.76\rpm0.66}$   \\
 & $\mathcal{L}_{photo}$($\times10^{\text{\scriptsize -3}}$) &$8.04\rpm4.46^*$&$5.75\rpm3.37^*$&$5.54\rpm3.31^*$&$8.83\rpm3.48^*$&$8.86\rpm4.14^*$&$\mathbf{4.86\rpm2.53}$  \\
 & $\perthousand$ of $|J_{\phi}| \leq 0$ & $\mathbf{0.01\rpm0.03^*}$ &$0.04\rpm0.04^*$ &$0.51\rpm0.32^*$ &$2.35\rpm0.99^*$ &$2.71\rpm6.71^*$ &$0.14\rpm0.05$    \\ \\
 
\multirow{6}{*}{$R=8$} 
& SSIM  &$0.78\rpm0.04^*$ &$0.79\rpm0.05^*$ &$0.78\rpm0.05^*$ &$0.79\rpm0.05^*$ &$0.78\rpm0.12^*$ &$\mathbf{0.80\rpm0.07}$   \\
 & PSNR &$37.27\rpm2.95^*$ &$37.28\rpm3.27^*$ &$37.69\rpm3.44^*$ &$35.42\rpm2.80^*$ &$34.07\rpm4.58^*$ &$\mathbf{38.65\rpm2.90}$   \\
 & NRMSE &$0.20\rpm0.05$ &$0.21\rpm0.08^*$ &$0.20\rpm0.07^*$ &$0.23\rpm0.08^*$ &$0.29\rpm0.13^*$ &$\mathbf{0.19\rpm0.11}$  \\
 & HFEN  ($\times 10^{\text{\scriptsize -2}}$) &$1.96\rpm0.48^*$ &$2.12\rpm0.71^*$ &$2.27\rpm0.68^*$ &$2.57\rpm0.55^*$ &$2.97\rpm1.12^*$ &$\mathbf{1.68\rpm0.38}$  \\
 & $\mathcal{L}_{photo}$($\times10^{\text{\scriptsize -3}}$) &$9.21\rpm3.93^*$&$5.85\rpm3.15^*$&$6.96\rpm2.14^*$&$9.17\rpm2.73^*$&$7.96\rpm3.55^*$&$\mathbf{5.11\rpm2.21}$  \\
 & $\perthousand$ of $|J_{\phi}| \leq 0$ &$\mathbf{0.01\rpm0.01^*}$ &$0.02\rpm0.02$ &$1.21\rpm0.51^*$ &$4.56\rpm1.54^*$ &$1.73\rpm3.27^*$ &$0.03\rpm0.01$   \\ \\
 
\multirow{6}{*}{$R=16$} 
& SSIM &$0.77\rpm0.04^*$ &$0.77\rpm0.05^*$ &$0.77\rpm0.05^*$ &$0.76\rpm0.05^*$ &$0.74\rpm0.12^*$ &$\mathbf{0.79\rpm0.05}$\\
 & PSNR &$37.08\rpm2.92^*$ &$37.21\rpm3.27^*$ &$36.21\rpm3.29^*$ &$34.13\rpm2.75^*$ &$34.20\rpm4.46^*$ &$\mathbf{38.58\rpm3.08}$  \\
 & NRMSE &$0.24\rpm0.05$ &$0.29\rpm0.08^*$ &$0.31\rpm0.07^*$ &$0.34\rpm0.08^*$ &$0.38\rpm0.13^*$ &$\mathbf{0.21\rpm0.06}$   \\
 & HFEN  ($\times 10^{\text{\scriptsize -2}}$) &$2.87\rpm0.45^*$ &$3.03\rpm0.74^*$ &$3.41\rpm0.69^*$ &$3.14\rpm0.61^*$ &$3.24\rpm1.78^*$ &$\mathbf{2.38\rpm0.41}$    \\
 & $\mathcal{L}_{photo}$($\times10^{\text{\scriptsize -3}}$) &$9.27\rpm3.97^*$&$5.96\rpm2.97^*$&$8.69\rpm2.69^*$&$9.38\rpm2.78^*$&$8.02\rpm3.41^*$&$\mathbf{5.26\rpm2.28}$  \\
 & $\perthousand$ of $|J_{\phi}| \leq 0$ &$0.08\rpm0.14^*$ &$\mathbf{0.01\rpm0.01^*}$ &$3.18\rpm1.24^*$ &$4.69\rpm1.69^*$ &$1.49\rpm2.79^*$ &$0.04\rpm0.04$   \\ \hline
\end{tabular}
\begin{tablenotes}
 \item * For p-value $<0.05$.
\end{tablenotes}
\end{threeparttable}
\end{table*}

\begin{figure*}[h!]
\centerline{\includegraphics[width=\textwidth]{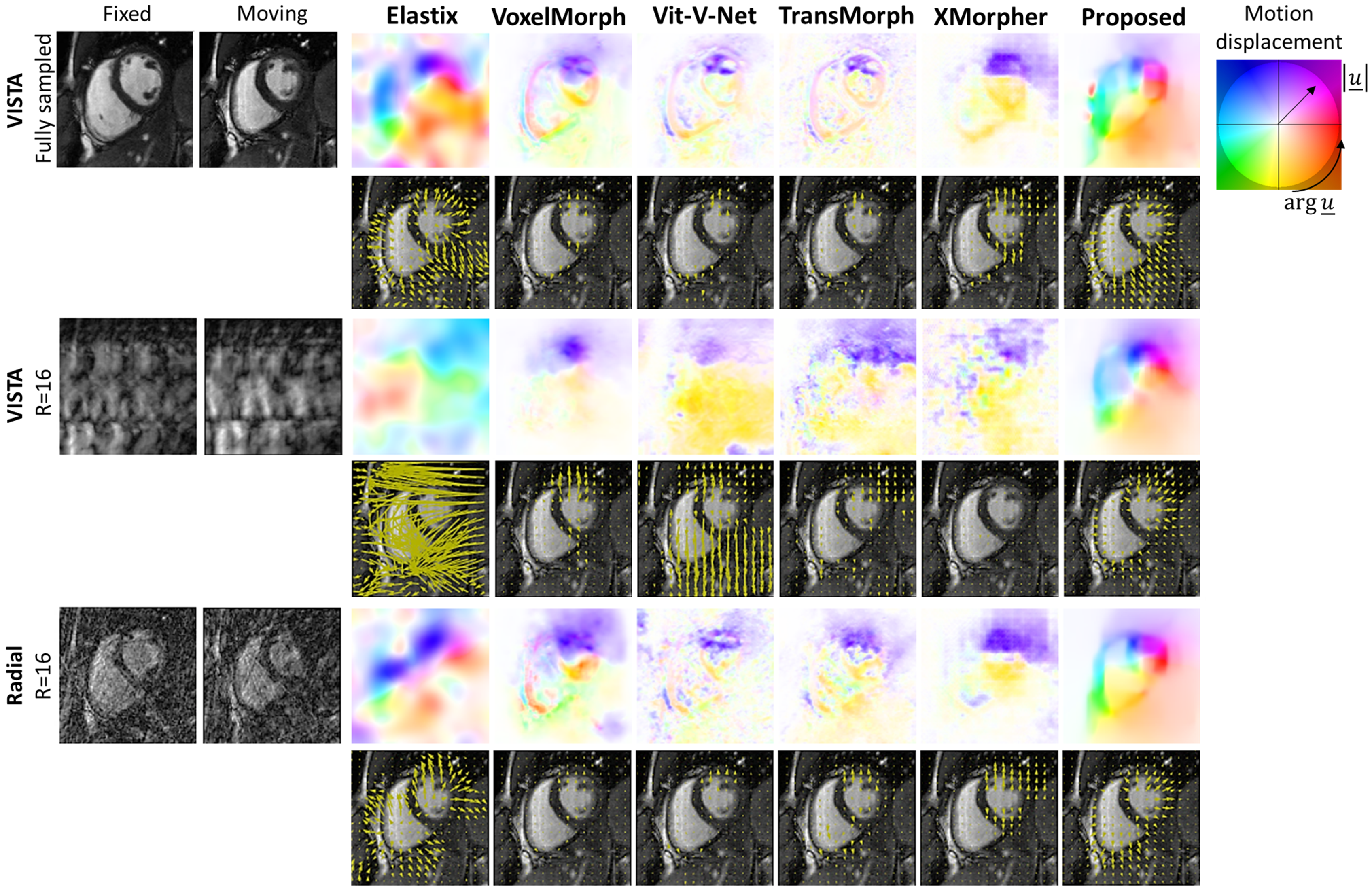}}
\caption{Cardiac motion estimation by the proposed framework compared to Elastix\cite{klein2009elastix}, VoxelMorph\cite{balakrishnan2019voxelmorph}, Vit-V-Net\cite{chen2021vit}, TransMorph\cite{chen2022transmorph} and XMorpher\cite{shi2022xmorpher} in a healthy subject. Predicted flow fields are shown for the fully sampled and retrospectively undersampled acquisitions with VISTA (Cartesian) and radial (non-Cartesian) undersampling for $R\!=\!16$ acceleration. To demonstrate the estimated motion direction and amplitude, motion estimations are color-encoded following \cite{baker2011database} and in the second row of panels depicted as quiver plot overlays on the fully sampled moving image. Our model demonstrated consistent and superior performance compared to other methods.}
\label{flow_estimation}
\end{figure*}

We studied the registration performance of the models based on the residual photometric error as described in \eqref{eq:photo_loss} between the transformed images, obtained by warping the fully sampled moving image with the estimated motion, and the corresponding fully sampled fixed image. We also reported the individual Dice scores (DSC) \cite{dice1945measures} for the right ventricle (RV) and the left ventricle (LV) masks obtained by registering the end-systolic masks to the end-diastolic masks and the end-diastolic masks to the end-systolic masks. The ground truth masks were automatically determined with the Segment software \cite{heiberg2010design} and then corrected manually. The regularity of the estimated motion fields was estimated using the percentages of non-positive values in the determinant of their Jacobian matrix ($\%$ of $|J_{\phi}| \leq 0$). Additionally, we analyzed the influence of high accelerations on motion estimation accuracy and robustness by looking at the color-encoded motion estimates \cite{baker2011database}.

\subsection{Motion-compensated Reconstruction Evaluation}

We collected motion estimates for each frame with respect to all other frames within the motion cycle, i.e. each time a different fixed image is taken from the motion cycle. These estimations are then used to reconstruct the complete image volume as described in section \ref{sec:recon}. The number of neighboring frames used for the motion-compensated reconstruction was investigated, i.e. indicating the capture of long-range dependencies and the ability to estimate occluded motion between far-apart frames. We varied the number of neighboring frames $\rpm T$ to instill different amounts of temporal information in the reconstruction process between $\rpm 2$ and $\rpm 12$. Motion-compensated reconstructions were compared qualitatively and quantitatively to the reconstructed images from the fully sampled k-space data. To study the reconstruction performance, we used the Structural Similarity Index (SSIM) for image quality perception\cite{Zhou2004image}, the Peak Signal-to-Noise Ratio (PSNR) to assess the recovery of the image details, the Normalized Root Mean Square Error (NRMSE) to evaluate the differences between the two images and the High-Frequency Error Norm (HFEN) to quantify the quality of the fine features and edges\cite{ravishankar2010mr}. The mean and the standard deviation of the averaged metrics across all slices from the test subjects were reported. The statistical significance of differences between the studied methods and the proposed framework was assessed using a paired $t$-test with Bonferroni correction.

\subsection{Ablation Experiments}

We performed a set of ablation studies to investigate the effect of the building blocks of our model on the registration performance. All ablated versions were trained on the cardiac dataset. We ablated on cardiac motion estimation for the fully sampled case and using the Cartesian VISTA undersampling with accelerations $R\!=\!8$ and $R\!=\!16$. Different variants of the model were trained that either keep or remove the examined component in isolation. We ablated over the GMA module, the update operator (GRU, LSTM\cite{shi2015convolutional}) and the denoiser (no denoiser, Warm-starting, ResNet denoiser, UNet denoiser). The UNet denoiser has three levels. Each level has two convolutional layers with each having $3\times3$ filters with $1$-stride followed by a rectified linear unit. The number of feature maps starts at $32$ and doubles at each stage. For warm-starting, their input images were first reconstructed using conjugate-gradient SENSE\cite{pruessmann2001advances} to reduce the undersampling artifacts and provide improved image quality and then fed to the proposed network. For each ablated model, we used grid search to find improved hyperparameters for the training loss parameters, the learning rate and the weight decay, with the aim of minimizing the residual photometric error averaged over selected validation subjects from the test data. 

\section{Results}
\subsection{Motion Estimation Results}
\subsubsection{Cardiac Motion Estimation}
\begin{figure}[h]
\centerline{\includegraphics[width=\columnwidth]{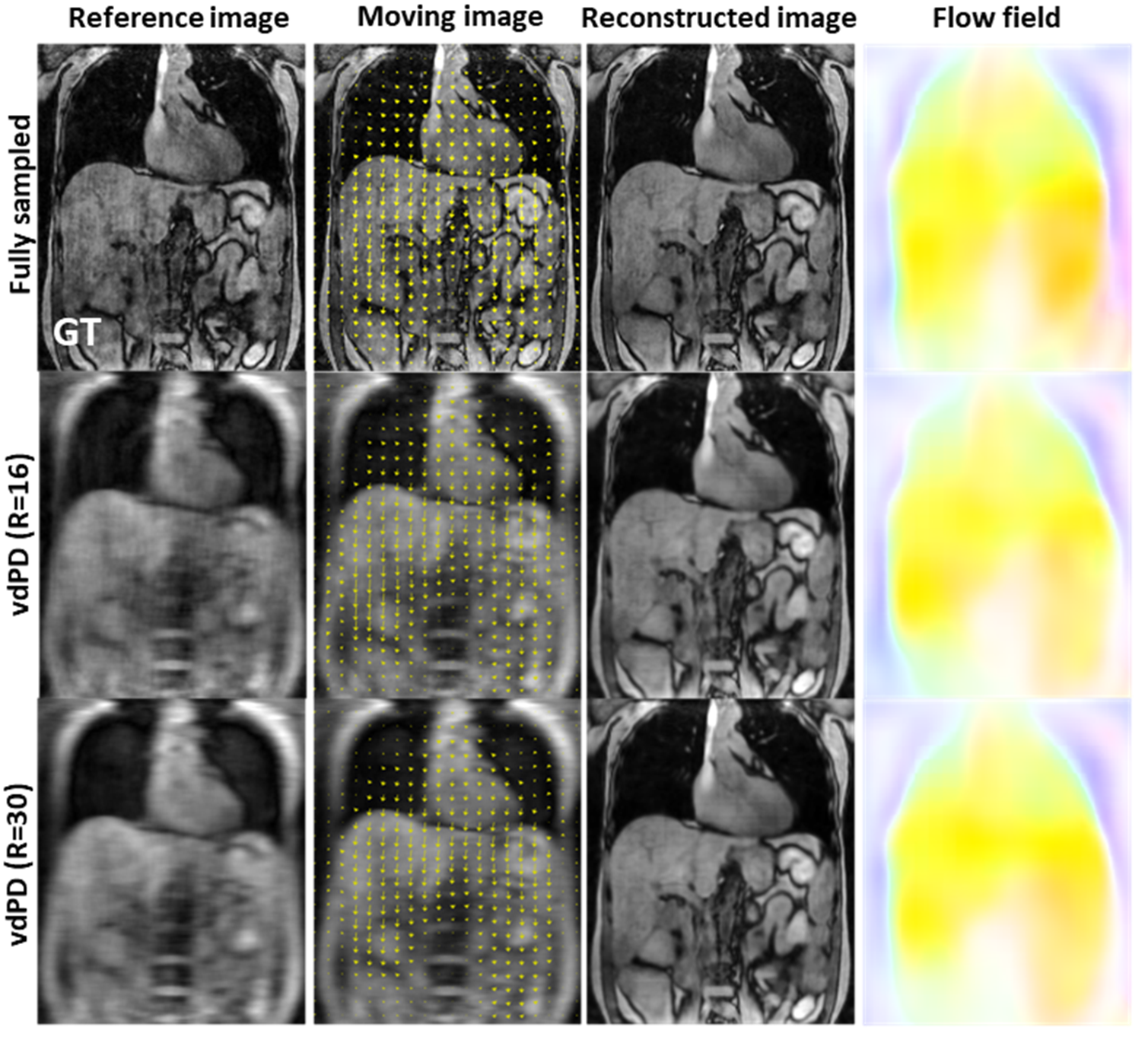}}
\caption{Respiratory motion estimation in a patient with neuroendocrine tumor and unconfirmed liver metastasis for the fully sampled and Cartesian vdPD accelerated acquisitions with accelerations $R\!=\!16$ and $R\!=\!30$. Deformation
fields are overlaid on the moving image. Images of motion-compensated reconstructions are depicted next to the used color-encoded deformation fields indicating motion from end-expiratory to end-inspiratory state. Consistent performance over different accelerations was observed.}
\label{resp}
\end{figure}
Figure \ref{DSC} shows the Dice measures for the right and the left ventricles as violin distributions comparing the proposed method to other registrations. Our model achieved the highest mean Dice scores for the Cartesian and radial trajectories for the fully sampled and $R\!=\!8$ and $R\!=\!16$ accelerations and outperformed the comparative approaches for both masks (p-value $<0.01$) with fewer outliers. Like the Dice score findings, the residual photometric error, reported in Table \ref{tab_comp_vista}, indicated that the proposed model produced more accurate registration outcomes and remained consistent over Cartesian and non-Cartesian acquisitions at high accelerations with statistical significance (p-value $<0.001$). The percentage of pixels with non-positive Jacobian determinant is presented in Table \ref{tab_comp_vista}. The regularity measure was higher in the transformer and the hybrid competing networks (p-value $<0.01$) which indicates that the smoothness constraint in this case is not enough to enforce more regular and diffeomorphic transforms. Unlike these methods, the percentage of pixels exhibiting folding (i.e., \% pixels with $|J_{\phi}| \leq 0$) in the motion estimates provided by our model decreased with higher accelerations reflecting the smoothing effect despite the presence of artifacts. Although VoxelMorph (Cartesian and radial) and Elastix (radial) displacements exhibited nearly no folding, we found that enforcing more regularity in our model resulted in oversmoothing the boundaries and hence the loss of structural and local details. 

Qualitative comparisons across all registrations are presented in Figure \ref{flow_estimation} which showcases the effect of the sampling trajectory with high accelerations on the motion estimation in a healthy subject. The color-encoded motion estimations and the corresponding quiver plots overlaid on the fully sampled moving image are displayed. The motion estimates of our method concentrated motion towards the cardiac region with a static background and a closer agreement to the underlying relaxation motion. Accurate motion directions for the different anatomical components were obtained while separating well the moving parts from the background and the left and the right ventricular boundaries for the fully sampled and $R\!=\!16$ accelerated estimates for the two sampling trajectories. While our motion estimation differs from those of other learning-based approaches, it has the closest alignment with the results produced by the state-of-the-art conventional baseline, Elastix, in the fully sampled case. Nevertheless, Elastix exhibits excessive spatial smoothing along the background and the boundaries of anatomical structures. This oversmoothing results in a loss of edge and fine-grained information that is essential for local measures of segmentation overlaps and the subsequent task of motion-compensated reconstruction. Elastix failed to locate the heart for the VISTA undersampling and delivered unsatisfactory results. On the other hand, other deep learning methods reduced the relaxation motion of the left ventricle to a translation. Considerable ambiguity was observed in terms of boundary recognition of the right ventricle. VoxelMorph yielded smooth results despite high acceleration. Contrarily, discontinuities characterized the estimates of XMorpher, the full transformer model. The hybrid networks that combine convolutional networks and transformers (Vit-V-Net and TransMorph) showed local disruptions too. Transmorph delivered for the VISTA undersampling noisy estimates with unrealistic motion forms that followed the undersampling artifacts.

\subsubsection{Respiratory Motion Estimation}

Figure \ref{resp} depicts the performance of the proposed approach in detecting respiratory motion in a patient with a neuroendocrine tumor and unconfirmed liver metastasis. We illustrate the color-encoded deformation fields and the results of motion-compensated reconstruction for the fully sampled case and using the Cartesian vdPD undersampling with acceleration rates $R=16$ and $R=30$. Consistent performance was observed throughout increasing acceleration factors. No outlier or failed misregistrations were found. The motion estimates showed prominent superior-inferior respiratory motion with a smoothed background. High-quality motion-compensated reconstructions were obtained showing good agreement with the reference. Edges like the diaphragm or liver vessels were well preserved for the fully sampled and the accelerated cases.

\subsection{Comparison of Motion-compensated Reconstructions}

The influence of the number of neighboring frames considered for the motion-compensated reconstruction on the reconstruction performance is reported in Figure \ref{neighbouring_frames} and Table \ref{tab_neighbouring}. We compared the reconstructed image quality and the corresponding error maps in systole obtained with VoxelMorph and our model using the Cartesian VISTA undersampling with $R\!=\!16$ for a patient with suspected right ventricular cardiomyopathy. The appearance of anatomical structures deteriorated as the considered number of neighboring frames increased. This indicated that frame-to-frame misregistrations accumulated throughout the cycle, particularly for a larger number of neighboring frames. Residual ghost and blurring artifacts remained present if the number of frames during reconstruction was kept small as insufficient information was available to reconstruct certain cardiac phases. Reconstructions obtained with the motion fields of the proposed model were less obstructed by the induced irregularities and misalignment from motion information as it showed good results even if all the frames were included in the reconstruction. This indicates that long-range frame dependencies have been learned. We chose to use $T=\rpm6$ for the reconstruction of each cardiac phase as the image quality remained most similar to the reference, with most of the motion artifacts being eliminated. This was also confirmed by the quantitative scores. Both models yield improved SSIM and PSNR for $T=\rpm6$ compared to other numbers of considered neighboring frames.

\begin{table}[h]
\centering

\begin{threeparttable}
\caption{Mean and standard deviation of SSIM and PSNR metrics for motion-compensated reconstruction using the proposed model and VoxelMorph motion estimation for $R\!=\!16$ Cartesian VISTA acceleration with variable neighboring frames $T$}
\label{tab_neighbouring}
\begin{tabular}{lllll}
 \hline
\multicolumn{1}{c}{\multirow{2}{*}{$T$}} & \multicolumn{2}{c}{VoxelMorph} & \multicolumn{2}{c}{Proposed}  \\
 & \multicolumn{1}{c}{SSIM} & \multicolumn{1}{c}{PSNR}  & \multicolumn{1}{c}{SSIM} & \multicolumn{1}{c}{PSNR} \\
 \hline
 $\rpm2$  & $0.64\rpm0.06^*$ & $33.16\rpm2.51^*$ & $0.63\rpm0.06$ & $32.60\rpm2.60$ \\

 $\rpm4$  & $0.69\rpm0.06^*$ & $35.47\rpm2.90^*$  &
$0.71\rpm0.06$ & $35.40\rpm2.90$\\

 $\rpm6$  &
$0.72\rpm0.06^*$ & $35.65\rpm3.06^*$ &
$\mathbf{0.74\rpm0.07}$ & $\mathbf{36.04\rpm3.08}$ \\

 $\rpm8$  &
$0.72\rpm0.06^*$ & $35.52\rpm3.08^*$ &
$0.73\rpm0.06$ & $35.80\rpm3.10$ \\
 $\rpm12$ &
$0.71\rpm0.06^*$ & $35.25\rpm3.06^*$ &
$0.73\rpm0.06$ & $35.60\rpm3.10$\\
\hline
\end{tabular}
\begin{tablenotes}
 \item * For p-value $<0.05$.
\end{tablenotes}
\end{threeparttable}
\\

\vspace{12pt}

\includegraphics[width=\columnwidth]{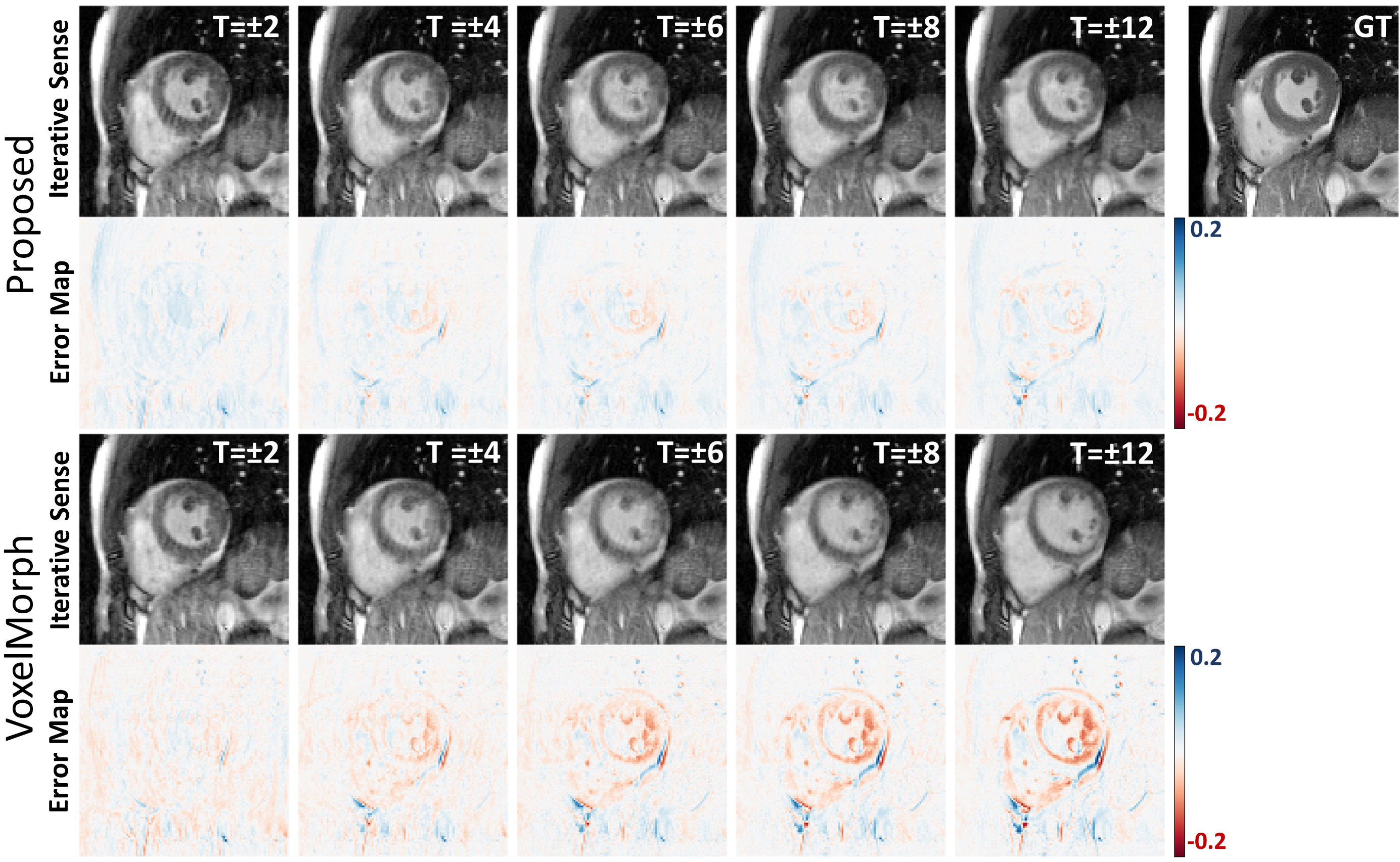}
\\
\captionof{figure}{Motion-compensated reconstructions and respective error maps between the fully sampled reference image and reconstructions using different neighboring frames $T$ in a healthy subject. Motion estimations were obtained from Cartesian VISTA accelerated acquisitions with $R\!=\!16$ using the proposed model (top) and VoxelMorph (bottom).}
\label{neighbouring_frames}
\end{table}

\begin{figure*}[h]
\centerline{\includegraphics[width=\textwidth]{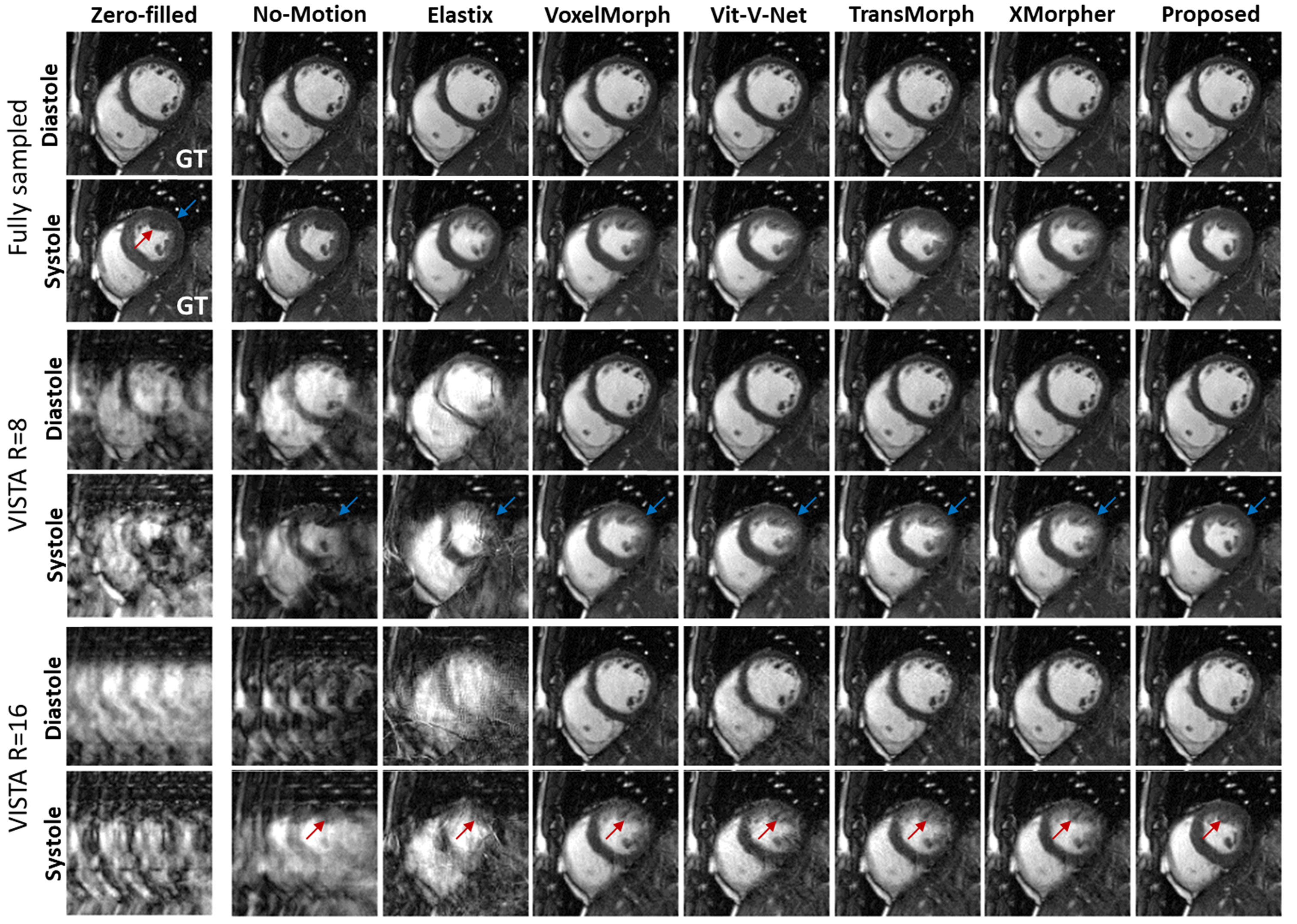}}
\caption{Motion-compensated reconstructions of the end-systolic and end-diastolic phase are shown for the fully sampled case and Cartesian VISTA acceleration of $R\!=\!8$ and $R\!=\!16$ in a healthy subject. Images were reconstructed using iterative Sense with no motion, Elastix\cite{klein2009elastix}, VoxelMorph\cite{balakrishnan2019voxelmorph}, Vit-V-Net\cite{chen2021vit}, TransMorph\cite{chen2022transmorph}, XMorpher\cite{shi2022xmorpher} and the proposed model motion estimates in comparison to the fully sampled reference image. The proposed method permitted the recovery of fine details like the papillary muscles (red arrows) while preserving the morphological structures and myocardial thickness (blue arrows) in systole despite high undersampling factors.}
\label{exp2_vista}
\end{figure*}

\begin{figure*}[h]
\centerline{\includegraphics[width=\textwidth]{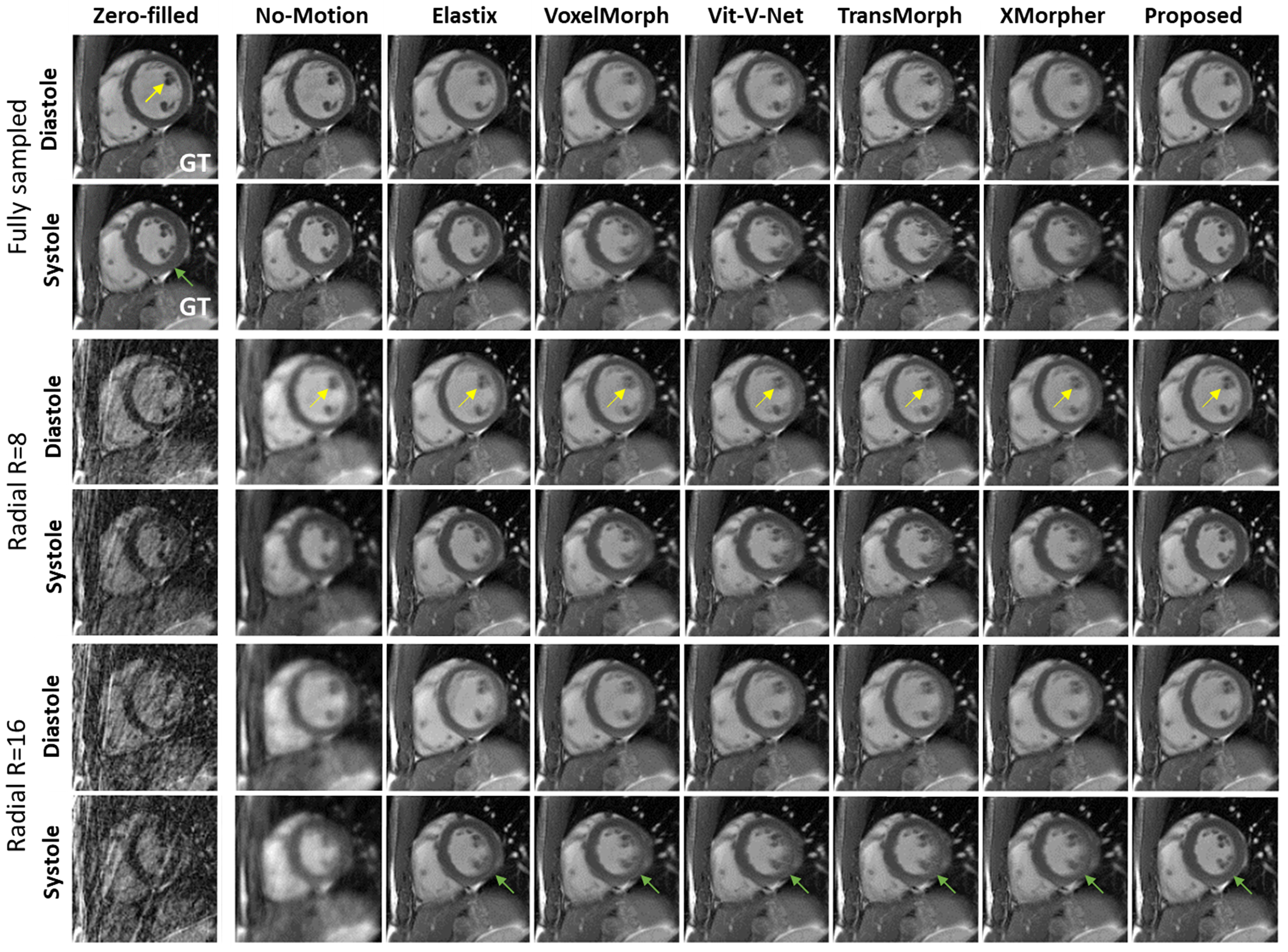}}
\caption{Motion-compensated reconstructions of the end-systolic and end-diastolic phase are shown for the fully sampled case and the accelerations $R\!=\!8$ and $R\!=\!16$ with a radial (non-Cartesian) trajectory in a patient with hypertrophic cardiomyopathy. Images were reconstructed using iterative Sense with no motion, Elastix\cite{klein2009elastix}, VoxelMorph\cite{balakrishnan2019voxelmorph}, Vit-V-Net\cite{chen2021vit}, TransMorph\cite{chen2022transmorph}, XMorpher\cite{shi2022xmorpher} and the proposed model motion estimates in comparison to the fully sampled reference image. Our approach yielded more accurate recovery to the target reference over all accelerations compared to other registrations. Yellow arrows point to the observed blur and loss of high-frequency information in the papillary muscles, while green arrows indicate the impaired lateral epicardial sharpness in the competing methods.}
\label{exp2_radial}
\end{figure*}

Table \ref{tab_comp_vista} tabulates the quantitative evaluation of the motion-compensated reconstruction in terms of SSIM, PSNR, NRMSE and HFEN. Over the two studied trajectories, our method had superior image quality metrics on all acceleration rates and outperformed other methods in terms of the HFEN metric (p-value $ < 0.01$) indicating the ability to better preserve high-frequency information. The reconstruction was improved with our method compared to Elastix in terms of SSIM and PSNR (p-value $ < 0.01$). Although Elastix performed poorly for the Cartesian undersampling, it yielded better image quality for the radial undersampling compared to the Cartesian one with comparable NRMSE to the results obtained by our method. The difference between the reconstructions obtained using the proposed motion estimates compared to those obtained with Vit-V-Net, TransMorph and XMorpher was statically significant for all the studied metrics in the fully sampled and accelerated cases of both trajectories (p-value $ < 0.01$). Superior reconstruction quality was obtained using the proposed model compared to VoxelMorph in the radial trajectory (p-value $ < 0.01$) and highly accelerated VISTA undersampling (p-value $ < 0.05$). For mild VISTA undersampling ($R\!=\!8$), the performance of both models remained comparable in terms of SSIM, NRMSE and PSNR.

The motion-compensated reconstruction of the proposed framework compared to other registrations is shown in Figure \ref{exp2_vista} for the fully sampled and undersampled acquisitions with undersampling factors $R\!=\!8$ and $R\!=\!16$ using the VISTA mask in diastole and systole of a healthy subject. The reconstruction without motion compensation was included to appreciate the motion correction influence. Additionally, we added zero-filling reconstructions to show the effect of undersampling artifacts on the input images. Motion-resolved reconstructions with our model achieved an apparent image quality improvement for the accelerated cases compared to the non-corrected acquisitions. They showed finer detail retention and better agreement with the reference compared to the other methods for the fully sampled case and highly accelerated acquisitions despite the presence of undersampling artifacts. Blurring of edges, motion underestimation and residual artifacts were observed in the results of other approaches. Elastix introduced apparent artifacts in its undersampled reconstructions. Most errors of prior deep learning methods were present in the systolic frames and became more prominent as the acceleration factor increased. They reduced the underlying non-rigid contraction/relaxation motion to a translation motion. The blue arrows showcase the blur in the epicardial and endocardial left ventricle wall during systole and the papillary muscles, marked by the red arrows, were not well depicted. 

Similar observations were made for the radial trajectory, as depicted in Figure \ref{exp2_radial}. The motion-compensated reconstructions for the fully sampled case and accelerations of $R=8$ and $R=16$ are shown for a patient with suspected hypertrophic cardiomyopathy. Reconstruction with motion estimation provided by the proposed method reflected the underlying cardiac dynamics, recovered more faithfully all the anatomical structures and showed sharper boundaries around dynamic regions. Myocardial wall thickness remained consistent with the reference and high-frequency details were better conserved throughout the tested accelerations compared to other studied methods. Elastix succeeded in resolving the motion pattern correctly even for the accelerated cases, yet motion underestimation was still observed in systolic frames. Meanwhile, reconstruction using motion estimates of other deep learning-based methods deformed the actual shape of the ventricles and showed blurring at image edges. Both systolic and diastolic frames became blurrier as the acceleration increased. This was especially apparent in the left ventricle boundaries and myocardial wall (green arrows) and the smaller structures such as the papillary muscles (yellow arrows).

\subsection{Ablation Studies}

Table \ref{tab_ablation} depicts the results of the ablation study. The GMA module improved the quantitative metrics for the accelerated cases. Yet, the model performance was still deteriorating as the undersampling factor increased. The denoiser network reduced this error and ensured a better performance for RAFT and the proposed model. Modest differences in image quality metrics were noted quantitatively between using ResNet (proposed) or UNet as a denoiser. However, ResNet provided better Dice scores and reconstruction quality on highly accelerated data and suppressed the artifacts more effectively. Using LSTM instead of GRU as an update operator exhibited performance fluctuations, which was also noted in previous studies \cite{teed2020raft, jiang2021learning} and did not yield any notable enhancement in image quality. Likewise, warm-starting had minimal impact on performance and introduced abrupt local discrepancies in motion estimations, impacting the model's ability to handle undersampling artifacts. These ablations highlight the role of the denoiser and the GMA modules in improving performance in the presence of aliasing caused by the undersampling.

\begin{table}[h!]
\caption{Quantitative comparison for ablation studies}
\label{table}
\centering
\begin{threeparttable}
\begin{tabular}{p{0.8cm} p{1.6cm} p{1.6cm} p{1.4cm} p{1.4cm}}
 \hline
\multicolumn{2}{l}{\textbf{Cartesian VISTA}} & Fully sampled& $R=8$ & $R=16$ \\
 \hline
\multirow{5}{*}{RAFT}	&
\parbox[t]{2cm}{SSIM\\ PSNR \\ NRMSE \\ HFEN ($ \times10^{\text{\scriptsize -2}}$) \\ DSC} &
\parbox[t]{2cm}{$0.81\rpm0.06^*$ \\ $38.47\rpm3.12^*$ \\ $0.21\rpm0.10^*$ \\ $2.31\rpm0.80^*$  \\ $0.73\rpm0.15^*$\\}&
\parbox[t]{2cm}{$0.69\rpm0.05^*$ \\ $36.37\rpm3.23^*$ \\ $0.30\rpm0.05^*$ \\ $2.22\rpm0.78$  \\ $0.68\rpm0.17^*$\\} & 
\parbox[t]{2cm}{$0.73\rpm0.05$ \\ $34.32\rpm3.12^*$ \\ $0.26\rpm0.06^*$ \\ $3.80\rpm0.9^*$  \\ $0.66\rpm0.17^*$\\}\\

\multirow{5}{*}{\begin{tabular}[c]{@{}l@{}}RAFT $+$\\ResNet \\ denoiser\end{tabular}}   & 
\parbox[t]{2cm}{SSIM\\ PSNR \\ NRMSE \\ HFEN ($ \times10^{\text{\scriptsize -2}}$) \\ DSC} &
\parbox[t]{2cm}{$0.81\rpm0.05^*$ \\ $39.33\rpm3.36^*$ \\ $0.19\rpm0.08^*$ \\ $2.18\rpm0.74^*$  \\$0.78\rpm0.18^*$ \\ }&
\parbox[t]{2cm}{$0.76\rpm0.06^*$ \\ $36.69\rpm3.31^*$ \\ $0.24\rpm0.05^*$ \\ $2.43\rpm0.78^*$  \\ $0.77\rpm0.17^*$\\} & 
\parbox[t]{2cm}{$0.72\rpm0.07^*$ \\ $35.25\rpm3.25^*$ \\ $0.27\rpm0.05^*$ \\ $2.94\rpm0.89^*$  \\ $0.70\rpm0.16^*$\\}\\

\multirow{5}{*}{\begin{tabular}[c]{@{}l@{}}Proposed \\(no \\ denoiser)\end{tabular}}   & 
\parbox[t]{2cm}{SSIM\\ PSNR \\ NRMSE \\ HFEN ($ \times10^{\text{\scriptsize -2}}$) \\ DSC}&
\parbox[t]{2cm}{$\mathbf{0.85\rpm0.04}$ \\ $40.05\rpm3.72$ \\ $0.17\rpm0.06$ \\ $2.02\rpm0.97$   \\ $0.81\rpm0.15$\\} &
\parbox[t]{2cm}{$0.76\rpm0.06^*$ \\ $36.84\rpm2.99^*$ \\ $0.23\rpm0.07^*$ \\ $2.68\rpm0.84^*$  \\ $0.78\rpm0.14^*$\\}  &
\parbox[t]{2cm}{$0.72\rpm0.07^*$ \\ $35.29\rpm3.03^*$ \\ $0.25\rpm0.07^*$ \\ $3.14\rpm1.2^*$  \\  $0.76\rpm0.17^*$\\} \\

\multirow{5}{*}{\begin{tabular}[c]{@{}l@{}}Proposed\\ (UNet\\denoiser)\end{tabular}} 	&
\parbox[t]{2cm}{SSIM\\ PSNR \\ NRMSE \\ HFEN ($ \times10^{\text{\scriptsize -2}}$) \\ DSC }	&
\parbox[t]{2cm}{$\mathbf{0.85\rpm0.04}$ \\ $40.39\rpm3.58$ \\ $\mathbf{0.17\rpm0.06}$ \\ $\mathbf{1.81\rpm0.84}$    \\ $0.82\rpm0.17$\\}& 
\parbox[t]{2cm}{$\mathbf{0.79\rpm0.02}$ \\ $37.52\rpm3.20$ \\ $\mathbf{0.20\rpm0.06}$ \\ $2.24\rpm0.46$  \\ $0.82\rpm0.17$\\} &
\parbox[t]{2cm}{$\mathbf{0.80\rpm0.05}$ \\ $35.58\rpm3.11$ \\ $\mathbf{0.23\rpm0.06}$ \\ $2.72\rpm0.87^*$  \\ $0.79\rpm0.15^*$\\} \\

\multirow{5}{*}{\begin{tabular}[c]{@{}l@{}}Proposed \\ (LSTM \\ update)  \end{tabular}}   & 
\parbox[t]{2cm}{SSIM\\ PSNR \\ NRMSE \\ HFEN ($ \times10^{\text{\scriptsize -2}}$) \\ DSC}&
\parbox[t]{2cm}{$0.85\rpm0.06$ \\ $39.93\rpm5.08$ \\ $0.18\rpm0.11$ \\ $1.89\rpm1.1$   \\$0.81\rpm0.19^*$\\ } &
\parbox[t]{2cm}{$0.77\rpm0.08$ \\ $37.36\rpm3.47$ \\ $0.21\rpm0.05$ \\ $2.42\rpm0.71^*$  \\ $0.79\rpm0.18^*$\\}  &
\parbox[t]{2cm}{$0.73\rpm0.05$ \\ $35.53\rpm3.28$ \\ $0.28\rpm0.05^*$ \\ $2.63\rpm0.93^*$  \\ $0.78\rpm0.18^*$\\} \\

\multirow{5}{*}{\begin{tabular}[c]{@{}l@{}}Proposed \\ (Warm-\\ starting)  \end{tabular}}   & 
\parbox[t]{2cm}{SSIM\\ PSNR \\ NRMSE \\ HFEN ($ \times10^{\text{\scriptsize -2}}$) \\ DSC}&
\parbox[t]{2cm}{$0.85\rpm0.08$ \\ $40.71\rpm4.18$ \\ $0.19\rpm0.13$ \\ $1.87\rpm1.02$   \\$0.83\rpm0.21$\\ } &
\parbox[t]{2cm}{$0.77\rpm0.07$ \\ $37.61\rpm4.71$ \\ $0.22\rpm0.07$ \\ $2.25\rpm0.98$  \\ $0.82\rpm0.24$\\}  &
\parbox[t]{2cm}{$0.73\rpm0.06$ \\ $35.32\rpm3.82$ \\ $0.25\rpm0.08$ \\ $2.51\rpm0.89$  \\ $0.81\rpm0.23$\\} \\

\multirow{5}{*}{Proposed} 	    & 
\parbox[t]{2cm}{SSIM\\ PSNR \\ NRMSE \\ HFEN ($ \times10^{\text{\scriptsize -2}}$) \\ DSC} &
\parbox[t]{2cm}{$0.85\rpm0.05$ \\ $\mathbf{40.97\rpm3.65}$ \\ $0.18\rpm0.08$ \\ $1.87\rpm0.77$    \\ $\mathbf{0.83\rpm0.16}$\\} &
\parbox[t]{2cm}{$0.78\rpm0.06$ \\ $\mathbf{37.82\rpm3.19}$ \\ $0.21\rpm0.07$ \\ $\mathbf{2.23\rpm0.76}$  \\ $\mathbf{0.82\rpm0.15}$\\} & 
\parbox[t]{2cm}{$0.74\rpm0.07$ \\ $\mathbf{36.08\rpm3.08}$ \\ $0.23\rpm0.07$ \\ $\mathbf{2.48\rpm0.86}$  \\ $\mathbf{0.82\rpm0.15}$\\}\\[-0.9em]
 \hline
\end{tabular}
\label{tab_ablation}
\begin{tablenotes}
 \item * For p-value $<0.05$.
\end{tablenotes}
\end{threeparttable}
\end{table}

\section{Discussion}

In this paper, we introduced a novel network for fast non-rigid pairwise image registration from highly accelerated MRI data. The proposed self-supervised model estimates cardiac and respiratory motion accurately from Cartesian and radial acquisitions. We leverage the attention mechanism to enhance the local motion features with high-level information by identifying visually similar source pixels that belong to the same anatomical structure and have coherent or complementary flow patterns. We use recurrent decoding to iteratively regress the motion estimation residuals. Despite including iterative updates and the attention mechanism that are known to be expensive, we maintain an efficient and lightweight framework. 

A sequential approach of first reconstructing the image followed by an image registration maintains good performance under mild acceleration scenarios. Nonetheless, to shorten the scan duration, as the acceleration factor increases, the registration process encounters challenges associated with the presence of undersampling artifacts within the images. In such cases, performing image registration on data affected by undersampling artifacts is necessary to achieve faster acquisitions. Reaching higher acceleration factors while keeping good image quality relies on finding a robust registration network for undersampled data. Our fine-detailed estimation permits the efficient integration of data from neighboring phases into the reconstruction of each phase within the motion cycle. Furthermore, accurate motion estimation in images acquired through highly accelerated imaging will reduce substantially the number of breath-holds needed for whole-heart cine MRI while preserving high image fidelity. 

We compared our method against conventional and recent self-supervised deep learning-based registrations that have different architectural designs. We demonstrated that our network outperforms the other approaches in terms of precision and robustness as it offers fine-detailed motion estimates resistant to blur and artifacts in various undersampling trajectories and accelerations. Additionally, we showed the superior performance of the proposed method in the downstream task of motion-compensated reconstruction. We embedded the motion estimates in the reconstruction and revealed that our model preserves better morphological and functional geometry (sharper edges, finer details). Quantitative and qualitative consistency with the reference is upheld more consistently across different sampling trajectories compared to competing methods, including fully sampled and accelerated scenarios.

The ablation study on our framework core components showed the importance of the interplay of the different architectural units to achieve optimal performance. The denoiser results in improved results for the accelerated cases. Pretrained on its own, the denoiser learns to reconstruct the undersampled images rather than smoothing out the undersampling artifacts in a beneficial way for the registration task. In this case, the attention matrix learns self-similarities of reconstructed images that misalign with the aggregated motion features, as these encapsulate correspondences extracted from the noisy images. On the other hand, training the model from the outset to handle images affected by undersampling artifacts (without warm-starting) leads to more robust motion estimation. The results suggest that the smoothed output of the denoiser encourages learning more meaningful image representations in the subsequent modules. In the context branch, we derive features exclusively from the fixed image as its learned representations encompass boundary and shape information that aligns with the characteristics of the targeted motion estimate. Our network combines local and global features that encode simultaneously detailed anatomical information and high-level characteristics. Convolutional networks encode visual similarities effectively at a local level conserving highly detailed information. However, their small receptive field hinders detecting spatial relationships between distant pixels \cite{luo2016understanding}. Hence, the global cues are necessary to provide a more complete understanding of the images by capturing long-range context among distant regions to help differentiate between anatomies and to determine high-level dependencies. The attention mechanism in the GMA module increases the receptive field of our model without requiring heavy computations. On their own, global features encounter difficulties in representing small details leading to pixelated motion estimates and inaccurate local displacements\cite{raghu2021vision}. The association of contextual sensitivity with low-level information in the recurrent operator offers the network more freedom during iterative updates to choose which information is needed to get a reliable description of the motion. While relying on detailed information to resolve motion, global context helps to unravel uncertainties on boundaries or easily confused areas caused by blur and undersampling artifacts and eliminates outliers that affect the estimation. This is ensured by the gating mechanism in GRU cells that facilitates dynamic decision-making regarding the retention or discarding of information, offering the network adaptability to capture intricate patterns and relationships across the different decoded features \cite{gruber2020gru}.

We assume that motion is mainly in-plane in pairwise registration. As a result, misregistration errors during reconstruction can accumulate as the number of frames considered for motion-compensated reconstruction increases. We chose to reconstruct data from a portion of the available neighboring cardiac phases and thereby improve reconstruction quality. Using more cardiac phases results in smoother images and reduces residual noise. Nonetheless, accurate motion correction and the preservation of the captured cardiac function are likewise critical for the diagnostic value of the image.  Hence, we restrain the range of cardiac displacements by limiting the number of considered frames in the reconstruction.

In the future, we want to further extend our network architecture to perform groupwise registration on 3D volume images. Currently, we perform individual 2D pairwise image alignment. Including additional temporal and spatial information will help in accounting for through-plane and trough-frame motion, better accommodating to anatomical variations and noise-like undersampling artifacts and ultimately improving the overall processing accuracy. Although the proposed network has fewer FLOP compared to the other deep-learning approaches and fewer trainable parameters compared to hybrid and transformer-based models, it operates on 2D data. In contrast, the other investigated methods are designed to handle 3D data. A 3D extension of our method will result in a considerable increase in the computational overhead for training and testing. This can be managed using alternative cost volume modeling like correlation decomposition \cite{xu2021high} and sparse matches \cite{jiang2021learning2}. Furthermore, we aim to investigate the integration of the registration network within an end-to-end trained motion-compensated reconstruction framework. We acknowledge some limitations in this study. Our training database had a limited size, covered a restricted number of pathologies and cardiac functions variety and was retrospectively undersampled. Other collected data may have features that are not accounted for which may result in suboptimal performance. Our study demonstrated the effectiveness of our model in respiratory motion estimation, but additional experiments are needed for a comprehensive review. Moreover, more realistic reconstruction settings will be investigated in future studies instead of focusing only on magnitude coil combined processing. Our model tends to oversmooth motion on edges in the presence of undersampling artifacts which reduces motion estimation precision as the amount of artifacts increases. Encouraging the network to learn another auxiliary task like segmentation may improve boundary detection for accelerated acquisitions by leveraging more semantic context.

\section{Conclusion}
We proposed a self-supervised deep learning-based image registration network that combines local features and global self-similarities to provide a reliable estimate of the motion from fully sampled and highly accelerated acquisitions. Our approach is compatible with different motion types (cardiac and respiratory motion) and trajectory designs (Cartesian and radial). We demonstrated higher registration accuracy and more consistency compared to different state-of-the-art registrations. Incorporating the deformation fields into a motion-compensated reconstruction showed that the proposed method outperforms state-of-the-art conventional methods and recent deep-learning algorithms of different architecture designs.

\end{document}